\documentclass[aps,prl,reprint,superscriptaddress, longbibiography]{revtex4-1}

\usepackage{color}
\usepackage{soul}
\usepackage{dcolumn}
\usepackage{graphicx}
\usepackage{amsmath}
\usepackage{amssymb}
\usepackage{bm}

\begin{document}

\title{Superstrong coupling in circuit quantum electrodynamics}

\author{Roman Kuzmin}
\affiliation{Department of Physics, Joint Quantum Institute, and Center for Nanophysics and Advanced Materials,
	University of Maryland, College Park, MD 20742, USA.}

\author{Nitish Mehta}
\affiliation{Department of Physics, Joint Quantum Institute, and Center for Nanophysics and Advanced Materials,
	University of Maryland, College Park, MD 20742, USA.}

\author{Nicholas Grabon}
\affiliation{Department of Physics, Joint Quantum Institute, and Center for Nanophysics and Advanced Materials,
	University of Maryland, College Park, MD 20742, USA.}

\author{Ray Mencia}
\affiliation{Department of Physics, Joint Quantum Institute, and Center for Nanophysics and Advanced Materials,
	University of Maryland, College Park, MD 20742, USA.}

\author{Vladimir E. Manucharyan}
\affiliation{Department of Physics, Joint Quantum Institute, and Center for Nanophysics and Advanced Materials,
University of Maryland, College Park, MD 20742, USA.}

\date{\today}

\begin{abstract}

\end{abstract}


\begin{abstract}
    
Vacuum fluctuations fundamentally affect an atom by inducing a finite excited state lifetime along with a Lamb shift of its transition frequency~\cite{milonni2013quantum}. Here we report the reverse effect: modification of vacuum modes by a single atom in circuit quantum electrodynamics~\cite{schoelkopf2008wiring}. Our one-dimensional vacuum is a long section of a high wave impedance (comparable to resistance quantum) superconducting transmission line~\cite{kuzmin2018quantum}. It is directly wired to a transmon qubit circuit~\cite{koch2007charge}.
Owing to the combination of high impedance and galvanic connection, the transmon's spontaneous emission linewidth can greatly exceed the discrete transmission line modes spacing~\cite{Devoret2007How}. This condition defines a previously unexplored \textit{superstrong} coupling regime of quantum electrodynamics where many frequency-resolved vacuum modes hybridize with a single atom~\cite{meiser2006superstrong}. We establish this regime by observing the spontaneous emission line of the transmon, revealed through the mode-by-mode measurement of the vacuum's density of states. The linewidth as well as the atom-induced dispersive photon-photon interaction are accurately described by a physically transparent Caldeira-Leggett model, with the transmon's quartic non-linearity treated as a perturbation. Non-perturbative modification of vacuum, including inelastic scattering of single photons~\cite{goldstein2013inelastic}, can be enabled by the superstrong coupling regime upon replacing the transmon by more anharmonic qubits~\cite{garcia2008quantum, le2012kondo, PhysRevLett.111.163601, forn2017ultrastrong, gheeraert2018particle}, with broad implications for simulating quantum impurity models of many-body physics~\cite{gogolin2004bosonization, affleck2008quantum}.

\end{abstract}

\maketitle

\noindent\textbf{INTRODUCTION}\\
\\
Our device consists of a split Josephson junction terminating a two-wire transmission line, which itself is made of a linear chain of $4\times 10^4$  junctions (Fig.~1a,b). The opposite end of the line is connected to a dipole antenna for performing local RF-spectroscopy. Recently it was shown that the collective electromagnetic modes of such chains are microwave photons with a velocity as low as $v \approx 10^6~\textrm{m/s}$ and wave impedance as high as $Z_{\infty} \sim  R_Q$, where $R_Q = h/(2e)^2 \approx 6.5~\textrm{k}\Omega$ is the superconducting resistance quantum~\cite{kuzmin2018quantum}. This was a significant step forward following the earlier experiments on electromagnetic properties of short Josephson chains~\cite{Manucharyan113, Corlevi2006Duality, Hutter2011, masluk2012microwave,Bell2012superinductor, weissl2015kerr}. A propagating photon corresponds to a plane wave of voltage/current across/along the two chains, similarly to a coaxial cable. At low frequencies the dispersion is nearly linear and there is a band edge at about $\omega_p/2\pi \approx 20~\textrm{GHz}$ due to the Josephson plasma resonance of individual chain junctions. The split-junction differs from the rest once its Josephson energy $E_J(\Phi)$ and hence the plasma mode frequency $\omega_0 (\Phi)$ is detuned below $\omega_p$ by an external flux $\Phi$ through the loop. The anharmonicity of this mode is approximately given by the charging energy $E_C = e^2/2C_J$ of the split-junction's oxide capacitance $C_J$. For $E_C/h \approx 300~\textrm{MHz}$, a disconnected  split-junction is similar to a conventional transmon qubit~\cite{koch2007charge}. The electromagnetic vacuum seen by such a transmon is formed by the standing wave modes of the finite length $L$ transmission line.

\begin{figure}[b]
	\centering
	\includegraphics[width=\linewidth]{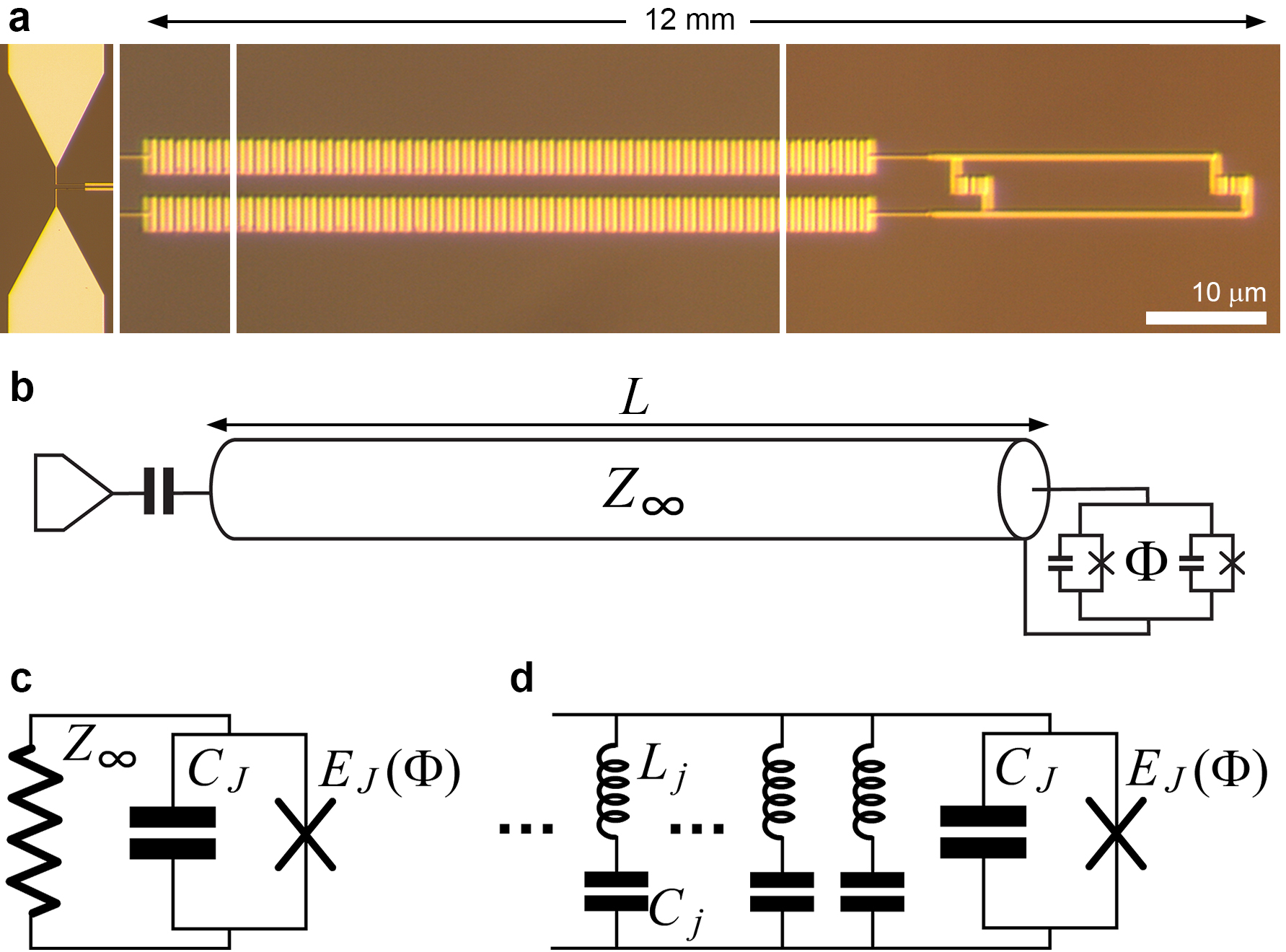}
	\caption{
		(a) Optical image of the device, showing (left to right) the coupling antenna, sections of the Josephson transmission line, and the split-junction termination. (b) Distributed circuit model of the device. The system is probed from the opposite to the split-junction end. (c) Equivalent circuit model of a semi-infinite transmission line with the wave impedance $Z_{\infty}$ shunting the split-junction. (d) A lumped element equivalent circuit model of (b) for a finite-length line.
		}
	
	\label{fig:Fig1}
\end{figure} 

Cavity QED typically describes a reversible exchange of an excitation between a single resonant mode and a qubit at the vacuum Rabi frequency $g$~\cite{haroche2006exploring}. For a long transmission line, the excitation exchange can occur faster than the time $2L/v$ that takes the photon to traverse the line. This would dramatically modify the mode wave function, e.g. replacing a node of electric field with an antinode, because to establish it a photon must complete the round trip. The fact that a single atom (qubit) can alter the photon's spatial profile motivated the term \textit{superstrong} coupling in the original proposal~\cite{meiser2006superstrong}. In the frequency domain, the condition for superstrong coupling is $g\rho > 1$, where $\rho = 2L/v$ is the density of vacuum modes ($1/\rho$ is the free spectral range), which means the qubit exchanges an excitation with multiple vacuum modes simultaneously. 
For a single mode system, another regime beyond strong coupling takes place at $g \sim \omega_0$, which was termed \textit{ultrastrong} coupling~\cite{forn2018ultrastrong,kockum2018ultrastrong}. 
In this regime the total number of excitations is no longer conserved, which leads to an entangled qubit-photon ground state~\cite{ciuti2005quantum}. From the many-body physics perspective, the most intriguing scenario is the combination of ultrastrong and superstrong regimes, where formation of multi-mode entanglement is expected~\cite{PhysRevLett.111.163601}.
In fact, quantum electrodynamics in this case maps onto the strongly-correlated (Kondo-like) quantum impurity models of condensed matter physics~\cite{garcia2008quantum, le2012kondo, goldstein2013inelastic, gheeraert2018particle}. The advantage of simulating such models in a quantum optics setting is that, for sufficiently large $g\rho$ and $g/\omega_0$, the many-body effects can be completely developed, while each microscopic mode remains individually accessible due to the finite system size. Moreover, the system size itself becomes an experimental control knob.

Spectroscopic signatures of ultrastrong coupling were previously reported in experiments on flux qubits and transmons
~\cite{Niemczyk2010, PhysRevLett.105.237001,bosman2017multi, yoshihara2017superconducting}. The superstrong coupling condition $g\rho > 1$ remains challenging even for circuit QED.
A pioneering experiment in this direction was done by capacitively coupling a transmon qubit to a nearly one-meter long coplanar waveguide transmission line with  $Z_{\infty} \approx 50~\Omega$~\cite{sundaresan2015beyond}. The conditions $g\rho \approx 0.3$ and $g \approx 30~\textrm{MHz}$  was achieved and multi-mode fluorescence was observed under a strong driving of the dressed qubit. However, the coupling was too low to expect quantum many-body effects with single photons. In a recent experiment the coupling was increased to $g\approx 160~\textrm{MHz}$ by using a higher impedance transmission line ($Z_{\infty} \approx 1.6~k\Omega$) made using Josephson junction chains~\cite{martinez2018probing}. Unfortunately, the mode density was also reduced due to the limited chain length, resulting in $g\rho \approx 0.4$. Moreover, increasing the coupling requires reducing the parameter $E_J/E_C$ of the transmon, which comes at the price of charge noise decoherence~\cite{jaako2016ultrastrong, manucharyan2017resilience}. Besides circuit QED, hybrid quantum acoustic systems are promising for creating vacua with a high density of discrete modes~\cite{moores2018cavity,han2016multimode}.

Remarkably, the coupling strength becomes practically unlimited with the galvanic connection used in our experiment ~\cite{Devoret2007How}. This is apparent in the $L \rightarrow \infty$ limit, where the line, as viewed by the transmon, can be replaced by an ideal resistance $Z_{\infty}$ (Fig.~\ref{fig:Fig1}c). The coupling is now characterized by the spontaneous emission rate $\Gamma$. On one hand (for a large but finite $L$) Fermi's golden rule gives $\Gamma = 2\pi g^2 \rho$~\cite{girvin2011circuit}. On the other hand, correspondence principle requires that the excited state lifetime $T_1 \equiv 1/2\pi\Gamma$ is given by a classical expression $T_1 =Z_{\infty} C_J$ (the ``RC" charge relaxation time). This gives $\Gamma = 2.5~\textrm{GHz}/Z_{\infty}[k\Omega]$, i.e. a typical transmon is overdamped by a conventional $50~\Omega$ vacuum. Same effect takes place with flux qubits~\cite{forn2017ultrastrong}. The main innovation of our experiment is to utilize a high wave impedance ($Z_{\infty}\approx 5-10~k\Omega$) in order to \textit{weaken} the galvanic coupling to the range $\Gamma \lesssim \omega_0/2\pi$ where the atomic transition is well-defined. The superstrong regime parameter $g\rho = \sqrt{\Gamma\rho/2\pi}$ can be adjusted independently of $\Gamma$ by varying the line length.\\

\noindent\textbf{RESULTS}\\
\\
In this work we achieved superstrong coupling with $g\rho \approx 1.3$ and we illustrate this by observing the phenomenon of spontaneous emission into a finite-size vacuum. As the parameter $g\rho$ grows above a unity, progressively more vacuum modes hybridize with the atom to the degree that the distinction between the two cannot be made. The spectrum no longer contains the individual $2g$-splittings, and the atomic transition is replaced by a band of about $\Gamma\rho$ states defining the spontaneous emission linewidth $\Gamma$. Clearly, reaching $\Gamma\rho \gg 1$ is needed for the many-body physics; in the two relevant experiments $\Gamma\rho \lesssim 1$~\cite{sundaresan2015beyond,martinez2018probing}. Here $\Gamma\rho = 2\pi (g\rho)^2 \approx 10$, which allowed us to observe, for the first time, the radiative broadening of an atom by measuring the frequency shifts that it induced on a large number of individual vacuum modes. 

\begin{figure*}
	\centering
	\includegraphics[width=\linewidth]{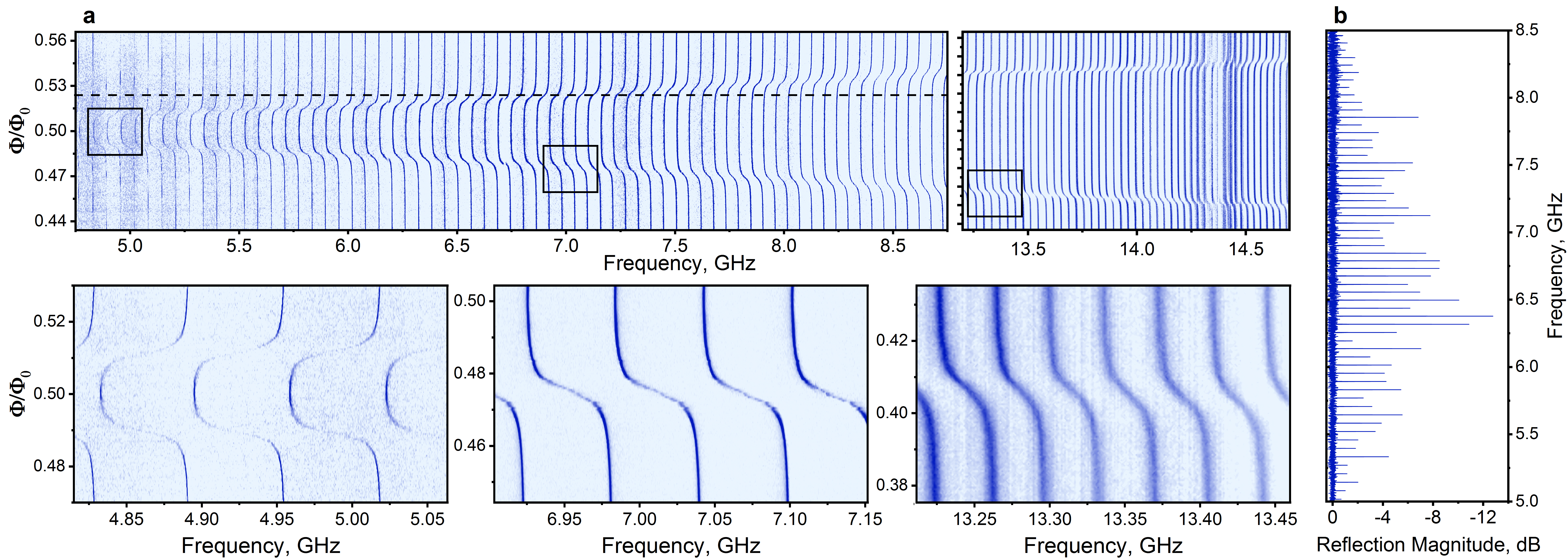}
	\caption{
		(a) Reflection magnitude as a function of drive frequency and flux through the loop. The color scale (not shown) is optimized to make the position of resonances maximally visible. The three insets have the same width along the frequency axis and correspond to the three boxed regions on the main plot. (b) An example slice of (a) at the external flux $0.525\times h/2e$, shown by the horizontal dashed line, where the atomic resonance is expected at around $7~\textrm{GHz}$.   		
	}
	
	\label{fig:Fig2}
\end{figure*}

We start with constructing a Hamiltonian model of galvanic light-matter coupling in our circuit. Accounting for the atom-induced distortions in the modes spatial profile can be challenging~\cite{bourassa2012josephson}. Consequently, even for $\Gamma \ll \omega_0$ the multi-mode Jaynes-Cummings model predicts divergent values of physical observables~\cite{PhysRevB.95.245115,PhysRevLett.119.073601}. Here we avoided these problems by choosing an appropriate circuit model of our system~(Fig.~\ref{fig:Fig1}d).   
An open section of a transmission line can be modeled by an infinite set of serial $L_jC_j$-circuits ($j =1,2,...$) connected in parallel. The frequency and characteristic impedance of each oscillator are given by  $\omega_j \equiv 1/(L_jC_j)^{1/2} = 2\pi(j-1/2)/\rho$ and $z_j \equiv (L_j/C_j)^{1/2}= Z_{\infty}\omega_j \rho/4$, respectively~\cite{devoret1995quantum}. A weak wave dispersion can be taken into account by introducing a slow frequency dependence to both $\rho$ and $Z_{\infty}$. We also introduce the bare transmon transition frequency $\omega_0 = ((8 E_J E_C)^{1/2}-E_C/2)/\hbar$ and linearized impedance $z_0 = R_Q(2 E_C/\pi^2 E_J)^{1/2}$. Defining the annihilation (creation) operators $a_0 (a_0^{\dagger})$ for the bare transmon and $a_j (a_j^{\dagger})$ for the uncoupled bath modes, the Hamiltonian reads:
\begin{equation}
\begin{aligned}
& H/\hbar = \omega_0 a_0^{\dagger}a_0 - K(a_0^{\dagger}a_0)^4+\sum_{j>0}\omega_j a_j^{\dagger}a_j -\\
&(a_0+a_0^{\dagger})\sum_{j>0} g_j(a_j +a_j^{\dagger}) +\left(\sum_{j>0} \frac{g_j}{\omega_0^{1/2}} (a_j + a_j^{\dagger})\right)^2,
\end{aligned} 
\end{equation}
where $K = E_C/2$ and individual couplings are given by $g_j = \omega_0(z_0/z_j)^{1/2}/2$~\cite{manucharyan2017resilience}. The vacuum Rabi frequency $g$ is given by $g = g_j (\omega_j \rightarrow \omega_0)/2\pi$ and we indeed confirm that $\Gamma = 1/2\pi Z_{\infty} C_J$.

The Hamiltonian (1) is a textbook Caldeira-Leggett model of a quantum degree of freedom interacting with an Ohmic bath~\cite{PhysRevLett.46.211}. The last term is the rigorously derived ``$A^2$"-term of quantum optics. Note that in our model $g_j^2 \sim 1/j$, which regularizes perturbative series automatically, without the need for sophisticated analysis recently proposed for a capacitively connected transmon~\cite{PhysRevB.95.245115, malekakhlagh2017cutoff}.  

Spectroscopy data as a function of drive frequency and flux through the split junction's loop is shown in Fig. 2. We observe a dense set of regularly-spaced resonances, with a mode spacing of about $1/\rho \approx 50-60~\textrm{MHz}$, which is disturbed periodically in flux (only a fraction of the period is shown). This flux-dependent disturbance is associated with the tuning of the trasmon's resonance across the spectrum of the transmission line. The unique feature of this data compared to previous multi-mode circuit experiments~\cite{sundaresan2015beyond,moores2018cavity} is that there are no individual vacuum Rabi splittings. In fact, at a fixed flux, the raw reflection trace vs. frequency contains no obvious sign of the atomic transition (Fig. 2b). However, upon tuning the flux, we observe that multiple resonances shift by nearly the same amount, indicating that many modes are equally sensitive to the tuning of the transmon (see three insets in Fig.2a). 

\begin{figure}
	\centering
	\includegraphics[width=\linewidth]{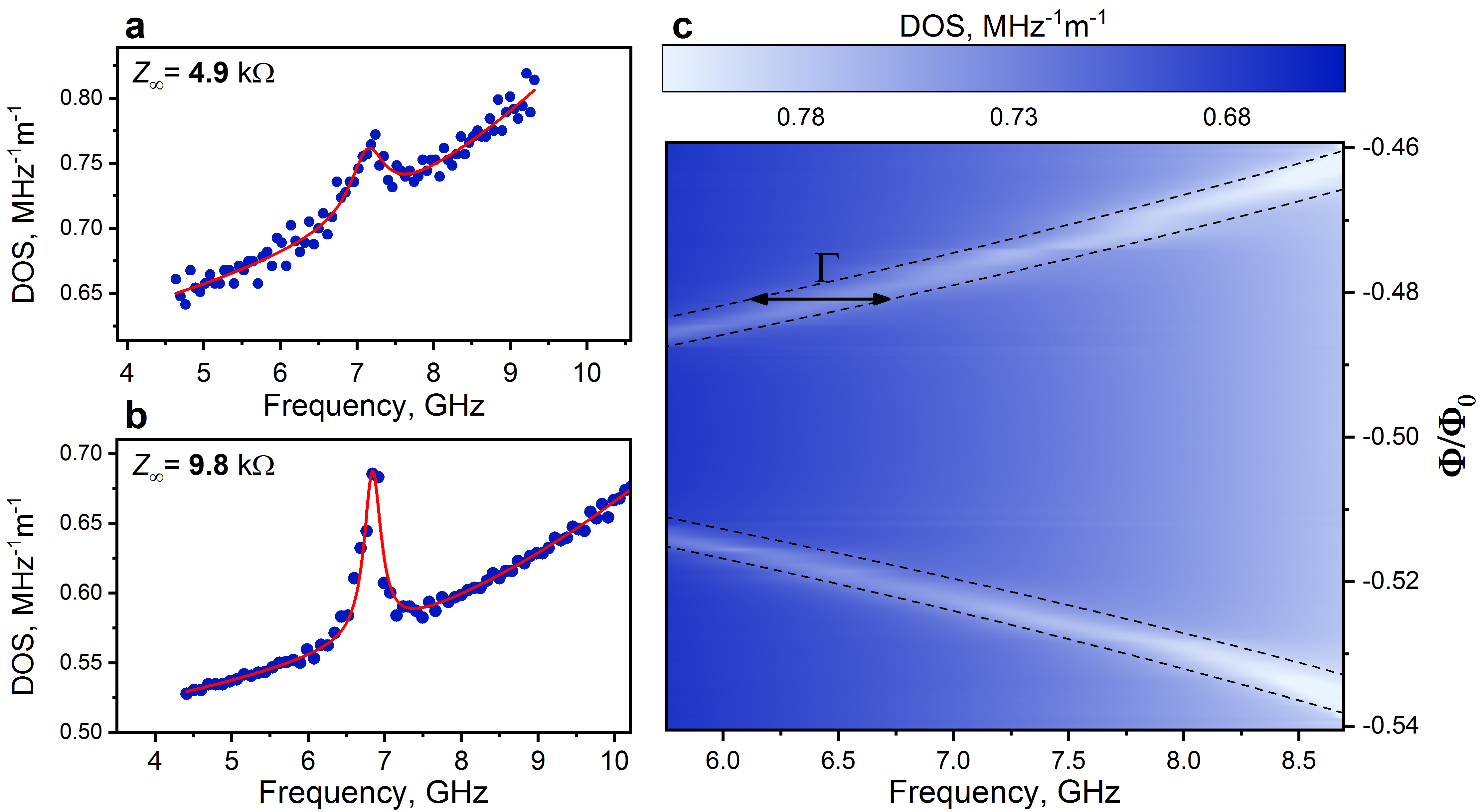}
	\caption{
		(a) Single-particle density of photon states (DOS) vs frequency (see text). The fluctuations are reproducible and reflect a small disorder in the junction parameters. The fit (solid line) is a combination of a Lorentzian resonance shape combined with the background transmission line DOS growing due to the van Hove singularity at the plasma frequency $\omega_p/2\pi = 22.6~\textrm{GHz}$. (b) DOS extracted for a device with $Z_{\infty} = 9.8~k\Omega$. (c) DOS from (a) plotted vs flux. The two dashed lines indicate the fit to a transmon's transition and a fixed width of the DOS peak.
	}
	
	\label{fig:Fig3}
\end{figure}

A clear picture of an atom simultaneously coupled to many vacuum modes comes from extracting the density of states (DOS) conventionally defined as $(\pi/L)/(\omega_{j+1}-\omega_j)$, where $\omega_j$ is the measured frequency of $j$th mode. This quantity is readily available since every individual mode is frequency-resolved in our reflection spectroscopy. The obtained DOS has a broad peak, involving about $20$ consequent modes, and a slowly increasing with frequency background (Fig. 3a). From fitting the background to a simple model of transmission line DOS, we extract the bare junctions plasma frequency $\omega_p/2\pi = 22.6~\textrm{GHz}$ and the wave impedance $Z_{\infty} = 4.9~k\Omega$ (Methods). The peak line-shape fits well to a simple Lorentzian function with unit area, which correctly reflects one additional state coming from the transmon. The DOS peak position shifts with flux and perfectly follows the disconnected transmon's resonance, assuming a reasonable junction asymmetry and that the maximal plasma frequency is that of the other chain junctions (Fig. 3c). Importantly, the fitted peak width is frequency-independent and matches the theoretical radiative linewidth $\Gamma \approx 1/2\pi Z_{\infty} C_J \approx 600~\textrm{MHz}$, within $\pm 50~\textrm{MHz}$. In a control experiment, we have checked that a device with a twice higher wave impedance,  $Z_{\infty} = 9.8~k\Omega$,  fabricated using a skinnier junction chain, narrows the DOS peak precisely by a factor of two (Fig. 3b).

To test the Hamiltonian (1) further, we measure the atom-induced  photon-photon interactions. Neglecting the transmon's non-linearity at first, the Hamiltonian (1) is quadratic and can be diagonalized exactly to obtain the flux-dependent normal mode frequencies $\tilde\omega_j (\Phi)$. The result very well reproduces the spectroscopy data in Fig.~\ref{fig:Fig2}. Expressing the Kerr term in terms of the ``quasiparticle" photon creation $c_j$ and annihilation $c_j^{\dagger}$ operators, and keeping only energy-conserving terms, we can approximate the Hamiltonian (1) as:  

\begin{equation*}
H/h \approx \sum_{j}\left( \tilde{\omega}_jc_j^{\dagger}c_j +  K_j(c_j^{\dagger}c_j)^2\right)  + \sum_{i\neq j}\chi_{i,j}(c^{\dagger}_i c_i c^{\dagger}_j c_j).
\end{equation*}
Here the $K_j$ is the induced Kerr shift of the $j$-mode frequency due to a single photon, which plays the role of the on-site repulsion in Bose-Hubbard models~\cite{jin2013photon}. The $\chi_{i,j}$ is the induced cross-Kerr shift of the mode $i$ due to a single photon in the mode $j$~\cite{weissl2015kerr} (Methods). 

We verified the presence of the induced Kerr effect using high-power spectroscopy. The modes inside the DOS peak clearly shift much more in response to a high-power driving compared to the modes outside of it (Fig. 4a). At some threshold power, we observe that the resonances within the DOS peak snap onto their uncoupled values. This effect, standard in circuit QED~\cite{bishop2010response}, can be also explained as the split-junction effectively acquires an infinite inductance because the swing of its overdriven phase exceeds $2\pi$.

\begin{figure}
	\centering
	\includegraphics[width=\linewidth]{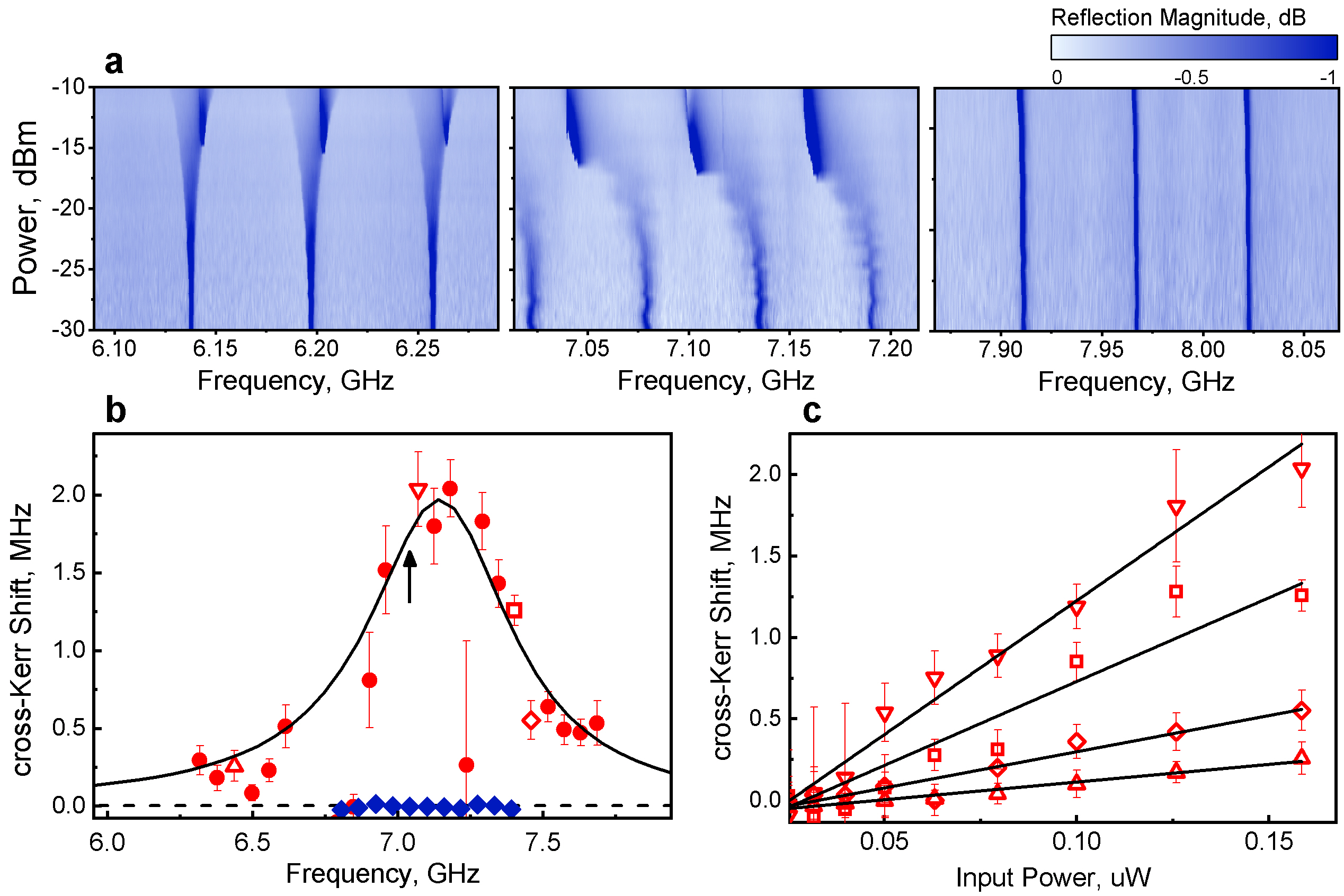}
	\caption{(a) Power-dependence of the spectroscopy signal for the DOS peak at $7.12~\textrm{GHz}$. The modes inside the DOS peak shift more strongly with power than the ones outside. (b) Dispersive shifts $\chi_{i,j}$ as a function of $i$-modes frequencies for a fixed second-tone driving of the mode $j=108$ at $7.014~\textrm{GHz}$, near the top of the DOS resonance (marked with an arrow). Blue markers indicate the same measurement for a largely detuned atom. Theory (solid and dashed lines) has no adjustable parameters except for the $Y$-axis scale (Methods). (c) Power dependence of the shifts of selected modes indicated by matching markers.}
	
	\label{fig:Fig4}
\end{figure} 

The cross-Kerr interaction can be characterized more accurately. For a chosen flux bias, we drive a mode $j =108$, corresponding to the peak in the DOS at a convenient atomic frequency around $7~\textrm{GHz}$. The power is such that the mode is populated by an approximately one photon on average. The neighboring modes $i=j\pm1, j\pm2, ...$ are scanned with a low-power drive to obtain the resulting  frequency shifts. This allows for a measurement of the dependence of $\chi_{i,j}$ on $i-j$ and comparison with our perturbative calculation. The $\chi_{i,j}$ is maximal for the DOS peak maximum, and drops rapidly as the $i$-modes leave the $\Gamma$-neighborhood of the atomic resonance. Although the measured shifts have some fluctuations of the presently unknown nature, the overall data matches theory well using the photon number to drive power conversion as the only adjustable parameter (Methods). \\

\noindent\textbf{DISCUSSION}\\
\\
The origin of a weak ($K \ll \Gamma$) photon-photon interaction strength is interesting and can be understood as follows. A mode of the bare transmon ($a_0$) is approximately an equal superposition of $\Gamma\rho \approx 10$ normal modes ($c_j$) of the combined system. Normalization requires that each mode amplitude is $1/(\Gamma\rho)^{1/2}$. Therefore, the induced Kerr and cross-Kerr shifts per photon due to the quartic non-linearity scale as $E_C/(\Gamma\rho)^2 \sim E_C/100$, which agrees with the data (Methods). Somewhat counterintuitively, increasing the light-matter coupling in the microscopic Hamiltonian (1) leads to the dissolution of the transmon's quartic non-linearity over a large number of linear modes, thereby suppressing the many-body effects.

Fortunately, non-linearity can be dramatically enhanced in our system by reducing the size of the split-junction such that $E_J/E_C \sim 1$. Because of large quantum fluctuations of the Josephson phase, enabled by $Z_{\infty} \sim R_Q$~\cite{Corlevi2006Duality}, the series expansion of the cosine non-linearity becomes invalid. Moreover, owing to the galvanic connection, the offset charge noise is eliminated by the large zero-frequency capacitance of the transmission line. In fact, for $E_J/E_C \sim 1$, our galvanic circuit QED system must be described by the boundary sine-Gordon quantum field theory. This quantum impurity model has a critical point at $Z_{\infty} = R_Q$~\cite{gogolin2004bosonization, affleck2008quantum}. It plays an important role in a broad range of phenomena, from dissipative quantum phase transitions~\cite{schmid1983diffusion,schon1990quantum}, to interacting electrons in one dimension~\cite{kane1992transmission}. Alternatively, the small junction can be replaced by either a flux~\cite{forn2017ultrastrong} or fluxonium~\cite{Manucharyan113, manucharyan2017resilience,Catelani2015Collective,Rastelli2015fluxonium} two-level-like qubits to implement the spin-$1/2$ Kondo impurity model in the bosonic representation.

In summary, we have demonstrated the operation of an ideal superconducting resistor, with a value comparable to resistance quantum $R_Q$, operating at frequencies beyond $10~\textrm{GHz}$, and whose internal modes are frequency resolved and hence are individually accessible. The spontaneous emission of a transmon qubit galvanically wired to it is not modified by the discreetness of the resistor spectral function at time scales shorter than the photon round trip time $\rho$, as long as the coupling to individual modes exceeds the mode spacing. Owing to the interaction which is strong already during time $\rho$, the dissipative dynamics of an artificial atom can be revealed by frequency shifts of the resistor's individual modes. This situation defines a novel \textit{superstrong} regime of light-matter coupling, relevant for exploring many-body physics with photons.

We thank Moshe Goldstein and Cristiano Ciuti for stimulating discussions and acknowledge funding from NSF, BSF, NSF-PFC, and ARO-NGC.\\
\\
\textbf{METHODS}\\
\\
\textbf{Fabrication and measurement.} The devices were fabricated on a highly resistive silicon substrate using the standard Dolan bridge technique. The process involves e-beam lithography on a MMA/PMMA bi-layer resist and the subsequent double-angle deposition of aluminum with the intermediate oxidation step. The spectroscopy measurements are done in a single-port reflection geometry using the wireless interface described in \cite{kuzmin2018quantum}. A global magnetic field was used to create a flux through the split junction loop.\\
\\
\textbf{The density of states.} The density of states in our system can be expressed as the sum of two contributions $$DOS(\omega)=\frac{1}{2\pi v(1-(\omega/\omega_p)^2)^{3/2}}+\frac{1}{2L}\frac{\Gamma}{(\omega-\omega_0)^2+{\pi}^2\Gamma^2}$$
The first term is the DOS for photons in a bare "telegraph" transmission line. It is easily obtained from their dispersion relation $\omega(k) = vk/\sqrt{1+(vk/\omega_p)^2}$. The second term takes into account one additional state coming from the transmon. It is expressed as a Lorentzian function normalized by the double transmission line length so the integration over the full frequency range and the length results in exactly one. By fitting the measured DOS to the above expression we found the transmon's resonance frequency $\omega_0$ and broadening $\Gamma$ as well as the photons velocity $v$ and cut-off frequency $\omega_p$.\\
\\
\textbf{Self- and cross-Kerr coefficients.} In order to find the expressions for $K_j$ and $\chi_{i,j}$ we start from the Hamiltonian (1) and limit the number of bath modes to $N = 1000$. Without the non-linear term the Hamiltonian (1) can be diagonalized with a unitary transformation given by a matrix $U$. $U$ is a block matrix, which ensures the commutation relations are preserved in the new basis $\bold{c}^{\dagger}, \bold{c}$.
$$\left(\begin{array}{l}\bold{a}\\\bold{a}^{\dagger}\end{array}\right)=\begin{pmatrix}
u&v\\-v&u
\end{pmatrix}\left(\begin{array}{l}\bold{c}\\\bold{c}^{\dagger}\end{array}\right)$$
Here $\bold{a}=(a_0,a_1,...,a_{N-1},a_N)^T$, $\bold{a}^{\dagger}=({a_0}^{\dagger},{a_1}^{\dagger},...,{a_{N-1}}^{\dagger},{a_N}^{\dagger})^T$, etc., and $u, v$ are $N+1$ by $N+1$ matrices. We can now express the quartic transmon's non-linearity $\sim(a_0-{a_0}^{\dagger})^4$ in the new basis. Counting the terms $({c_j}^{\dagger}c_j)^2$ and ${c_i}^{\dagger}c_i{c_j}^{\dagger}c_j$ we arrive at the expressions
$$K_j=\frac{1}{2}E_C({v_{1j}}^2-{u_{1j}}^2)^2$$
$$\chi_{i,j}=2E_C({v_{1i}}^2-{u_{1i}}^2)({v_{1j}}^2-{u_{1j}}^2)$$
We checked that the same results can be obtained in the perturbative calculation only starting from the linear circuit Hamiltonian \cite{weissl2015kerr}.\\
\\
\textbf{Measurement of the cross-Kerr shifts.} The dispersive cross-Kerr interaction of photons caused by the Josephson junction non-linearity appears as a mode frequency shift which is linear in the photon population of any other mode. We picked up the mode $j=108$ and changed its photon population with the second microwave tone. For every other mode in the vicinity of the DOS peak we performed a set of reflection measurements with the second tone being consecutively on or off. The mode frequency in every measurement was found from the fit of the reflection coefficient, both the real and imaginary parts, to the damped LC-oscillator model. The shift is then defined as an average mode frequency change between the on and off state of the second tone. The procedure was repeated for various second tone powers. To eliminate the detrimental effects of the flux jitter, we chose the shortest possible time between two consecutive measurements. Moreover the frequency of the second tone was modulated with a time scale much smaller than the measurement time and with an amplitude larger than the jitter amplitude. This allowed us to create a constant population of the driving mode disregarding its frequency fluctuation. We performed the same measurements in the same frequency range but with the transmon's resonance largely detuned. We found cross-Kerr shifts smaller by three orders of magnitude, which is consistent with our perturbative calculations of the bare transmission line non-linearity. This consistency made the rough photon calibration possible.





\providecommand{\noopsort}[1]{}\providecommand{\singleletter}[1]{#1}%

\end{document}